\documentclass[fleqn,twoside]{article}
\usepackage{espcrc2}

\usepackage{graphicx}
\usepackage[figuresright]{rotating}

\newcommand{\AmS}{{\protect\the\textfont2
  A\kern-.1667em\lower.5ex\hbox{M}\kern-.125emS}}

\title{Catching NGC~4051 in the low state with {\it XMM-Newton}}

\author{P. Uttley\address[soton]{School of Physics and Astronomy,
University of Southampton, \\ 
        Southampton SO17 1BJ, United Kingdom}%
        R. D. Taylor\addressmark,
        I. M. M$^{\rm c}$Hardy\addressmark,
        M. J. Page\address[mssl]{Mullard Space Science Laboratory,
University College London, \\
Holmbury St Mary, Dorking RH5 6NT, United Kingdom},
        K. O. Mason\addressmark,
        G. Lamer\address{Astrophysikalisches Institut Potsdam, An der
Sternwarte 16, \\
D-14482 Potsdam, Germany}
        and
        A. Fruscione\address{Harvard-Smithsonian Center for
Astrophysics, 60 Garden Street, \\
Cambridge, MA 02138, USA}}
       
\begin{document}

\begin{abstract}
The Narrow Line Seyfert~1 (NLS~1) galaxy NGC~4051
shows unusual low flux states, lasting several
months, when the 2-10 keV X-ray spectrum
becomes unusually hard (photon index$<1$) while the spectrum at lower
X-ray energies is dominated by a large soft excess.  A {\it Chandra}
TOO of the low state has shown that the soft excess and hard components are
variable and well-correlated.  The variability of the hard component
rules out an origin in a distant reflector.  Here we
present results from a recent {\it XMM-Newton} TOO of NGC4051 in the low
state, which allows a much more detailed examination of the nature of
the hard and soft spectral components in the low state.  We
demonstrate that the spectral shape in the low state is consistent
with the extrapolation of the spectral pivoting observed at higher
fluxes.  The {\it XMM-Newton} data also reveals the warm absorbing gas in emission, as
the drop in the primary continuum flux unmasks prominent emission
lines from a range of ion species.
\vspace{1pc}
\end{abstract}

\maketitle

\section{INTRODUCTION}
The NLS~1 galaxy NGC~4051 is highly variable in X-rays on both long
and short time-scales \cite{mch03}.  In particular, on long time-scales, the
X-ray light curve of NGC~4051 shows unusual low-flux states lasting
months, during which the source varies little (compared with the
normal variability amplitude) and
the 2-10~keV X-ray spectrum becomes extremely hard (photon index
$\Gamma\sim1$) \cite{gua98,utt99,utt03a}.  The first broadband X-ray
spectrum of a confirmed low state, obtained in 1998 by {\it BeppoSAX} \cite{gua98},
suggested that the hard spectral shape (above $\sim3$~keV) could be
explained as being due to pure reflection from distant cold material
(possibly the putative molecular torus).  At lower energies, a
different, much softer, spectral component was revealed
($\Gamma\sim3$) which, it was suggested, may be associated with extended
emission from hot gas on scales of hundreds of parsecs \cite{sin99}.
Thus the lack of variability in the low state spectrum, and its
unusual shape, were explained entirely in terms of emission and
reprocessing from large-scales, while the primary continuum was itself
`switched off'.  

Subsequently however, a {\it Chandra} observation of NGC~4051 in a
normal state in April 2000 \cite{col01} (and a later low state observation
in February 2001 \cite{utt03a}) showed that there is in fact no significant
soft X-ray emission on scales of hundreds of parsecs.  Furthermore, although the source spectrum
had the same unusual hybrid form of the 1998 low state (albeit at a
higher continuum flux level), {\it Chandra} showed that the continuum
in the low state is significantly variable in both hard and soft
bands, and both hard and soft components are well-correlated.  Thus,
{\it Chandra} showed that the unusual shape of the low state spectrum
was in fact intrinsic to the primary continuum emission and not due
entirely to reflection and emission on larger scales.  It
was suggested \cite{utt03a} that in the low state, the unusually hard spectral shape
seen above a few keV is simply due to an
extrapolation of the spectral variability of NGC~4051 seen at higher
fluxes \cite{tay03},
where the power-law continuum pivots, to become softer at
higher fluxes (and consequently much harder at low fluxes).
Furthermore, the {\it Chandra} 2001 low state observation showed
evidence for strong reflection, consistent with the presence of a reflection
component including a relativistic diskline, which is {\it
independent} of continuum flux (also observed in MCG--6-30-15
\cite{tay03,fab03}).  These results suggest that the continuum
emission processes in the low state are not
really distinct from those in the much briefer (hours) low-flux epochs
observed during the source's normal state, and therefore the physical
structure of the source, and perhaps the mode of accretion,
 is fairly stable despite the large-amplitude flux variability.

In order to test our earlier interpretation of the 2001 low state
data, which was somewhat limited by the relatively low throughput of
{\it Chandra}, we used a 33~ksec (useful exposure) 
{\it XMM-Newton} TOO to observe NGC~4051 in a low
state in November 2002 (the observation was triggered using our {\it RXTE}
monitoring campaign \cite{mch03}).  
In this paper, we present some of the
highlights of this observation and their relevance to our earlier
interpretation of the spectral shape and variability.  A more
detailed analysis can be found in forthcoming papers \cite{utt03b,pag03}.

\section{RESULTS}
\subsection{The flux-flux plot: evidence for spectral pivoting}
\begin{figure}[h]
\vspace{9pt}
\includegraphics[scale=0.45]{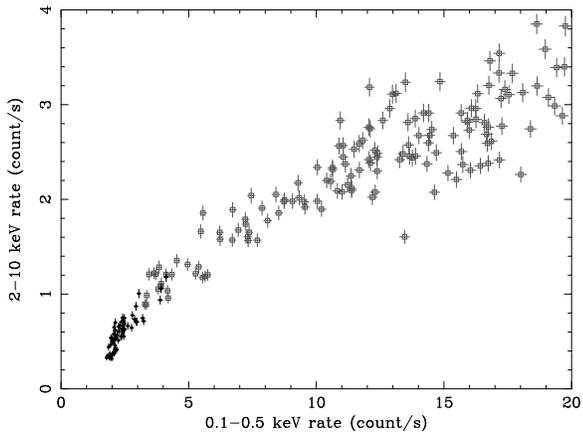}
\caption{Unbinned flux-flux relation.  Low state data
is plotted with filled markers, normal state data with lighter, open
markers.  For clarity only the 0-20 count/s soft flux range is
plotted, although all data (extending to 45 count/s) is included to
produced the binned relation (see Figure~\ref{fvfbin}).}
\label{fvf}
\end{figure}

In order to examine the form of the continuum spectral variability of
NGC~4051 in a model-independent way, and test whether the low state spectrum is consistent with
that expected from the spectral variability seen at higher fluxes, we
can produce a combined flux-flux plot \cite{tay03} of EPIC-pn fluxes obtained
during the low state TOO observation and an earlier (2001 May)
$\sim100$~ksec {\it XMM-Newton} observation of NGC~4051,
obtained during its normal, higher flux state.
The normal-state data is described in more detail
in \cite{mch03}.  For both low and normal states, we
obtained 500~s resolution light curves in the 0.1-0.5~keV soft and
2-10~keV hard bands. 
We first examine the unbinned flux-flux relation in Figure~\ref{fvf}.
We note two main points.  First, the overall
distribution of points clearly does not follow a linear relationship.
Second, the distribution of low state data seems to join smoothly on
to the distribution of normal state data: the normal state
distribution bends towards low fluxes and this bending continues in
the low state data.  Hence we conclude that the same process which causes
the spectral variability in the normal state continues to lower fluxes
in the low state.  There is significant intrinsic scatter in the
relation (which was also observed in the {\it RXTE} flux-flux
relations \cite{tay03}), which implies that there is
also spectral variability which is not correlated with flux
variations.  Note that this scatter increases towards higher fluxes,
which suggests that it is caused by some additional spectral variability in the pivoting
component, rather than independent variations in a separate component.

\begin{figure}[h]
\vspace{9pt}
\includegraphics[scale=0.45]{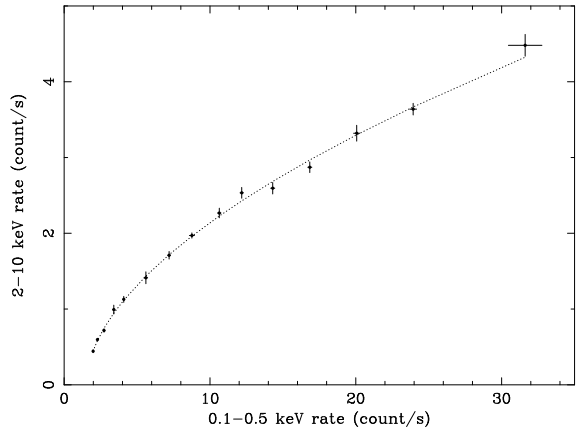}
\caption{Binned flux-flux relation.  The dotted line shows the best-fitting power-law
plus constants model described in the text.}
\label{fvfbin}
\end{figure}

To remove the scatter and so find the functional form of the
flux-flux relation, we binned
the data into 15 flux bins (minimum of 10 data points
per bin), which are logarithmically spaced in order to maximise the
usefulness of the low state data to constrain the flux-flux relation
(error bars are standard errors determined using the
spread of data points in each bin).  We use a general
power-law plus constants model \cite{tay03} to describe the
relationship of hard and soft fluxes, and obtain a good fit
to the data ($\chi^{2}=10.8$ for 11 d.o.f.) for a power-law index
$\alpha=0.57\pm^{0.07}_{0.04}$ with constant offsets on the hard and
soft axes of $C_{h}=0.1\pm^{0.25}_{0.1}$ and $C_{s}=1.62\pm^{0.3}_{0.07}$
respectively.  The binned flux-flux plot and best-fitting power-law
plus constants model are plotted in Figure~\ref{fvfbin}.
Using {\sc xspec} simulations we find that the 90\%
confidence range of power-law indices fitted to the binned
flux-flux plot corresponds to a pivoting model
with a pivot energy between 75-300~keV.  The positive offset on
the soft axis of the present flux-flux plot implies a significant constant
soft component which counteracts the effect of the hard constant
component detected at higher energies \cite{tay03} to produce an offset on the soft axis.

Since the spectral shape at low energies ($<1$~keV) is best described
by thermal emission \cite{utt03a,col01} rather than a power-law, then
in order to produce a flux-flux relation consistent with the pivoting
observed by {\it RXTE} \cite{tay03}, the 
pivoting power-law is probably attached to the thermal component and
pivoting may be driven by the variations in the thermal component.  An
obvious physical realisation of this picture is that the thermal
component (perhaps disk emission) is the source of seed photons which are Comptonised to
produce the power-law observed at higher energies.

\subsection{Evidence for reflection}
\begin{figure}[h]
\vspace{9pt}
\includegraphics[scale=0.45]{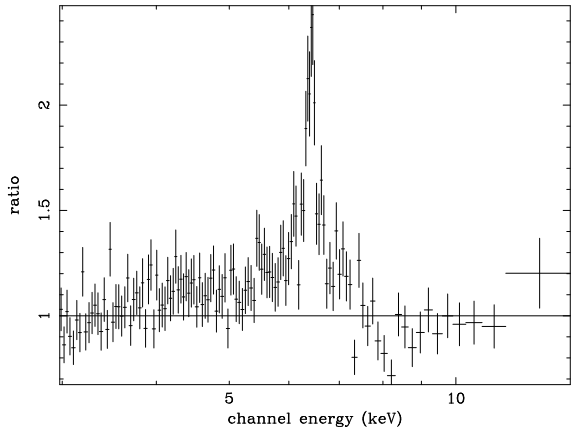}
\caption{Ratio of 3-12~keV EPIC-pn data to a simple power-law fitted
over that range but excluding data over 4-9~keV.}
\label{hardspec}
\end{figure}

In Figure~\ref{hardspec} we plot the
ratio of the 3-12~keV spectrum to a simple power-law model, which is fitted
only over the 3-4~keV and 9-12~keV ranges (i.e. excluding the region
where significant emission features may be expected).  Clear broad
residuals can be seen, in addition to a narrow iron emission line observed
at $\sim6.4$~keV.  Consequently,
a power-law plus narrow, unresolved Gaussian fitted to the data is a very poor fit
(reduced chi-squared, $\chi^{2}_{\nu}=1.94$ for 120 d.o.f.).
We find that including a {\sc pexrav}
reflection component (with parameters fixed to those used to describe the
May 1998 low state spectrum \cite{utt99})
improves the fit substantially
($R=2.4\pm^{0.5}_{0.7}$),
and an edge at $7.9\pm{0.15}$~keV (optical depth
$\tau=0.5\pm^{0.25}_{0.15}$)
is also formally required (resulting $\chi^{2}_{\nu}=0.91$ for 117
d.o.f).  The observed edge energy corresponds to the K-shell photoionisation
edge expected from Fe{\sc xvii}, a species which is seen in emission
at lower energies (see next section).
Removing the edge and including a Laor diskline, worsens the fit ($\Delta \chi^{2}=+13$ for
one less degree of freedom).  Adding the 7.9~keV edge in
addition to the diskline does not improve on the reflection+edge fit
significantly ($\Delta \chi^{2}=-3$ for three fewer degrees of
freedom).  Worse fits are obtained in each case by including a
diskline from a non-rotating black hole. 
Therefore we conclude that a diskline is not formally required to fit
the low state spectrum. However, the reflection required is factor$>2$
{\it larger}
than than that used to explain the entire May 1998 hard continuum
flux, while the narrow 6.4~keV line flux is {\it not larger} than observed
in May~1998.  Therefore we suggest that strong disk reflection is
in fact present in the low state spectrum, but the diskline may not be
detectable, perhaps due to extreme relativistic smearing in the
unstable inner disk.

\subsection{RGS spectrum}
\begin{figure}[h]
\vspace{9pt}
\includegraphics[scale=0.45]{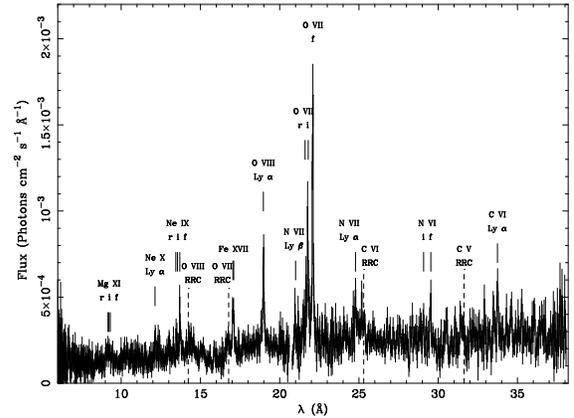}
\caption{{\it XMM-Newton} RGS spectrum of NGC~4051 in the low state.
Various prominent emission features are identified.}
\label{rgs}
\end{figure}

Finally, we show in Figure~\ref{rgs} the RGS spectrum obtained during
the 2002 low state observation.  The drop in the continuum from normal
flux levels reveals a host of emission lines, associated with the
ionised `warm absorber' gas, which has not had time to respond to the
drop into the low state.  The spectrum demonstrates how TOOs of AGN at
unusually low fluxes can clearly reveal their X-ray emission
lines, which may not be apparent in higher-flux observations where the
lines are more likely to be seen in absorption.  Besides providing
additional diagnostics in their own right \cite{pag03}, the emission
lines can be used to increase the information gained from grating
spectra obtained at higher fluxes, e.g. determining the true
optical depths of absorption lines which are partly filled in by
emission lines.

\end{document}